\RequirePackage{fix-cm}
\documentclass[smallextended]{svjour3}       
\smartqed  
\usepackage[bookmarks=true%
,bookmarksnumbered=true%
,hypertexnames=false%
,breaklinks=true%
,colorlinks=true%
,linkcolor=blue%
,citecolor=blue%
,urlcolor=blue%
]{hyperref}
\usepackage{graphicx}
\usepackage{epstopdf}
\usepackage{amsmath}
\usepackage{url}
\usepackage{graphicx}
 \usepackage{amssymb}
\usepackage{epstopdf}
\usepackage{epsfig}
\usepackage{mdwlist}
\usepackage{flushend}
\usepackage{array}
\usepackage{bm}
\usepackage{algorithm}
\usepackage{algpseudocode}
\usepackage{multirow}
\usepackage{colortbl}
\usepackage{balance}
\usepackage[T1]{fontenc}
\usepackage{tabularx}
\usepackage{tabulary}
\usepackage{refcount}
\usepackage{units}
\usepackage[table]{xcolor}
\usepackage[authoryear,round]{natbib}
\newcommand{\qn}{q_n}
\newcommand{\qr}{q_{n + 1}}

\newcommand{\added}{a_{n + 1}}

\newcommand{\snip}{s_{n}}
\journalname{Information Retrieval Journal}

\begin{document}

\title{A Term-Based Methodology for Query Reformulation Understanding}


\author{Marc Sloan         \and
        Hui Yang \and Jun Wang
}


\institute{M. Sloan \at
              University College London, UK \\
              \email{M.Sloan@cs.ucl.ac.uk}           
           \and
           H. Yang\at
              Georgetown University, USA \\ 
              \email{huiyang@cs.georgetown.edu} 
                         \and
           J. Wang\at
              University College London, UK\\ 
              \email{J.Wang@cs.ucl.ac.uk} 
}

\date{Received: 15 August 2014 / Accepted: 24 February 2015 / Published Online: 6 March 2015}

\maketitle

\begin{abstract}
Key to any research involving session search is the understanding of how a user's queries evolve throughout the session. When a user creates a query reformulation, he or she is consciously retaining terms from their original query, removing others and adding new terms. By measuring the similarity between queries we can make inferences on the user's information need and how successful their new query is likely to be. By identifying the origins of added terms we can infer the user's motivations and gain an understanding of their interactions.

In this paper we present a novel term-based methodology for understanding and interpreting query reformulation actions. We use TREC Session Track data to demonstrate how our technique is able to learn from query logs and we make use of click data to test user interaction behavior when reformulating queries. We identify and evaluate a range of term-based query reformulation strategies and show that our methods provide valuable insight into understanding query reformulation in session search. 
\end{abstract}


\section{Introduction}
\label{intro}

Session search in Information Retrieval (IR) occurs when a user issues multiple queries consecutively to a search engine in the pursuit of satisfying one or more information needs. A session is typically defined as a period of continuous interaction with a search engine and can be demarcated in a number of ways, a common one being 30 minutes of inactivity \citep{White:2007:IBV:1242572.1242576}. Sessions containing more than one query make up a significant proportion of search activity, with one study finding 32\% of sessions containing 3 or more queries \citep{Jansen:2005:TCA:1059467.1059470}. 
Understanding the underlying interactions in session search can lead to improved search interfaces, better search rankings and user satisfaction.

Sessions are driven by query reformulations, the user controlled act of modifying an existing query in order to pursue new search results. Query reformulations are usually closely related to the user's previous query and reflect the shifting cognition of the user throughout the session search. For instance, a user may have an unclear information need at the start of a session which becomes more refined as snippets are read and documents are clicked. Such queries can be ambiguous when the user is unsure how to explicitly define his or her information need \citep{Song:2009:IAQ:1508326.1508674} or explorative when the user is actively seeking a broad range of information on a subject \citep{Marchionini:2006:ESF:1121949.1121979}. In both cases, the information need can change throughout the session, whether through specialization, generalization and so on, which leads to variations in the queries used to describe it. 

We observe that sessions are typified by queries consisting of core terms related to the underlying information need and additional terms that reflect the user's cognitive changes \citep{Kinley:2012:HII:2414536.2414586}. Over the course of the session, the core terms may change as well. At any point in a session, we define three possible \emph{term actions} available to a user: 
\begin{description}
\item[\textbf{Term Retention}] - Keeping terms from one query to the next, the core terms for the current information need.
\item[\textbf{Term Removal}] - Removing a term from a query.
\item[\textbf{Term Addition}] - Adding a new term not present in the preceding query to the query reformulation. 
\end{description}

To illustrate a particular instance of query reformulation within session search and the described term actions, Table~\ref{trec-example} contains the queries in a typical search session found in the 2013 Session Track dataset \citep{TREC:2013:Session}. This session represents an explorative information need regarding public and political opinion on US gun control laws. The terms `{\tt gun control}' are \textbf{retained} through the first four queries, with the user \textbf{adding} and \textbf{removing} terms `{\tt opinions}', `{\tt US government}' and `{\tt current affairs}' in order to learn more about the topic. The focus shifts in query 5 with `{\tt gun control}' changing to `{\tt gun violence}', indicating a change in information need, which is expanded upon in the final query which is more specialized.

\begin{table} [t]\caption{Queries in session 40 of the TREC 2013 Session Track.}
\centering
\begin{tabular} {cl} 
\hline
Impression Position & Query \\
\hline
1 & {\tt gun control opinions} \\
2 & {\tt gun control us government} \\
3 & {\tt gun control current affairs} \\
4 & {\tt gun control current affairs} \\
5 & {\tt gun violence us} \\
6 & {\tt law center to prevent gun violence}\\
\hline
\end{tabular} \label{trec-example}
\end{table}

Without knowing the underlying information need driving the queries, the example demonstrates that it is possible to infer persistent subtopics and the terms that are likely to be retained or removed from query to query (in this case `{\tt gun control}' and `{\tt gun violence}'). A certain degree of overlap is typical between queries but how much? What factors influence whether a term is likely to be kept or removed in the next query? Can we determine a source for the new terms that are introduced into a query? Measuring the similarity between queries and other sources of text can help us resolve some of these questions and allow us to build descriptive and evaluative models of user behavior during a session search.

For instance, we observe in the example session that the snippets of all the results for the first query contain the terms `{\tt gun control}', and out of all ranked documents only the clicked document (ClueWeb ID \emph{clueweb12-0100wb-86-17546}) contains the terms `{\tt US government}' (in the phrase ``\emph{US Government Info Guide}''), which were then used in the next query. One inference that could be drawn here is that the user observed the terms `{\tt US government}' in the clicked document which influenced their reformulation decision making process. 

In this paper we seek to gain an understanding of the query reformulation process and resolve the following research questions:  
\begin{enumerate}
\item What is the relationship between terms found in adjacent queries in search sessions. How often are terms from a query retained or removed in a query reformulation?
\item Where are query reformulation terms not present in the original query sourced from and can we model term addition?
\item Can user-behavior scenarios defined on terms that are retained, removed or added inform us of the quality of query reformulations? 
\end{enumerate}
We resolve these questions by introducing a novel methodology for interpreting query reformulations using terms. We use our technique to explore term retention and removal by analyzing adjacent and non-adjacent queries in sessions. With term addition, our observations indicate that a significant number of added terms in a reformulation can be sourced from the terms that the user was exposed to in the previous impression. An impression consists of a query, its snippets and its documents, all of which contain terms that the user may have encountered during session search. By also incorporating click information, we can define and evaluate three sources for such terms, \emph{clicked} and \emph{non-clicked snippets} and \emph{clicked documents}.

The next stage in our analysis involved measuring the value of the three term sources in determining whether query terms were retained, removed or added, leading to eight possible scenarios of user behavior which we interpret based on our results. To evaluate the effectiveness of scenario-based term prediction, and also the user's observed query reformulations,  we determine whether the term actions ultimately lead to increased user satisfaction or improved search rankings, which we measure using implicit click information and a number of IR metrics. 

Our analysis was conducted on the TREC Session Track data from 2011 to 2014 \citep{TREC:2011:Session,TREC:2012:Session,TREC:2013:Session}, a set of standardized query logs comprising queries grouped by sessions across a number of predefined topics, the ranked documents, their snippets and clickthroughs (including order and dwell time) and relevance judgments. The documents belong to the ClueWeb09\footnote{\label{clueweb9}\url{http://www.lemurproject.org/clueweb09/index.php}} and ClueWeb12\footnote{\label{clueweb12}\url{http://www.lemurproject.org/clueweb12.php/}} corpora. This dataset was chosen as it is widely available, well regarded in the IR community and whilst small when compared to commercial query logs, is rich with potential sources for term discovery (snippets and documents), interaction data (clicks and dwell time) and relevance judgments (for evaluation).
		
The remainder of the paper is organized as follows. Section~\ref{related} presents the related work and Section~\ref{setup} outlines the dataset used, experimental setup and the key definitions and similarity measures used in our methodology. In Sections~\ref{retention} and \ref{discovery} we use our term-based technique to understand the three term actions \emph{retention}, \emph{removal} and \emph{addition}, investigate user click behavior and define the three term sources. In Section~\ref{experiment} we expand the term sources into user interaction based term scenarios and evaluate reformulation strategies. We conclude the paper and discuss our findings in Sections~\ref{discussion} and \ref{conclusion}. 

\section{Literature Review}
\label{related}

\emph{Session Log Analysis}\ \ \ Ours is not the first query log analysis of query reformulation behavior. \citet{Jansen:2009:PQR:1568763.1568772} defined different query reformulation states and the transition patterns that occur during a session and evaluated over a large query log. Their research idea is similar to our scenario approach although in their study the states operate on a query level by looking at the degree of overlap between queries, rather than our term based approach, but some of our findings are similar. \citet{ChangLiu:2010} explored a similar state-based analysis but this time on a user study that allowed them to determine different types of behavior based on the type of task being performed by the user. \citet{Kinley:2012:HII:2414536.2414586} also performed a user study with the intention of observing different query modifying behavior (such as replacing, adding terms etc.) and linking it to a user's `cognitive style' of query reformulation. A similar work to ours is Huang's \citep{Huang:2009:AEQ:1645953.1645966} classification of different types of reformulation behavior that utilizes clicks from query logs and uses term differences as well. Nonetheless, ours is the first such study using a purely term-based approach that also incorporates clicks in a user interaction model.

\emph{Click and User Modeling}\ \ \ A key component of this work is our click based methodology and our rank and impression position experiments. This is similar to work in click modeling, an established area of IR research that typically uses search logs, eye-tracking and user studies to understand how users navigate search pages. For instance, our $\snip(\mbox{LC})$ definition and experiments in Section~\ref{snippet-analysis} are based on the examination hypothesis model \citep{Joachims:2005:AIC:1076034.1076063,clickmodels}.  In other research, eye-tracking has been used on participants with predefined search tasks, with the researchers being able to predict which task was being performed based on eye tracking patterns \citep{Cole2011346,Cole:2010:LST:1962300.1962323}, which was further developed into being able to factor in the stage of the user's task \citep{Liu:2010:PIR:1835449.1835457}. Another recent eye tracking study \citep{Liu:2014:SRT:2661829.2661907} found that when browsing search results users will glance at snippets but not fully read them, returning to them at a later point if at all. These studies give in-depth insight into how users behave during search tasks which goes beyond what we model in this paper, although we too are interested in inferring user's reading and reformulation patterns.

\emph{Related Work}\ \ \ The work by \citet{Guan:2013:UQC:2484028.2484055} on session search re-ranking based on query and impression term matching is a similar approach to ours, although we build a more complex model to capture user interactions and we do not perform document re-ranking. Another similar work is by \citet{Jiang:2014:SBC:2600428.2609633} who conduct a comprehensive user and eye tracking study to understand how users behave over the course of a session. Their work includes statistics on reformulation behavior and ranking metrics across queries in sessions and many of their results mirror our own. Both pieces of research can be seen as a specialization of our methodology (for instance focusing on a particular type of term source) that concerns a specific IR problem, whereas ours is a more general study on trying to understand reformulation behavior. 

The work most similar to ours is the work by \citet{Liu201113847} on using terms from clicked snippets to aid in query recommendation. They recognize, as we do, that information needs persist through adjacent queries in search sessions but are difficult to define based purely on previous queries, and so use snippets as an additional term source. Unlike our methodology, they only use clicked snippets whereas we also incorporate terms from non-clicked snippets and documents, as well as the previous query. Where our work mainly diverges is that their objective is to locate terms that are useful for query recommendation, whereas our objective is to identify useful term sources for query reformulation (of which clicked snippets is one) under a number of conditions including clicks, rank and impression position. 

The work in this paper differs from the literature in that: 1) our methodology is term-based rather than query or task-based 2) our methodology is derived from data rather than a user or eye-tracking study and 3) our model incorporates clicks and differentiates term sources such as snippets and documents as sources of reformulation terms.

\section{Analytical Setup}
\label{setup}

\begin{table} [t]\caption{TREC 2011, 2012, 2013 and 2014 Session Track data overview.}
\centering
\begin{tabular} {r | c | c | c | c} 
\multicolumn{1}{ c |}{\multirow{2}{*}{ }} & \multicolumn{4}{c}{TREC Session Track} \\
 \cline{2-5}
  & \multicolumn{1}{ c |}{2011} & 2012 &  2013 & 2014\\
\hline
Number of topics & 62 & 48 & 49 & 51\\
Number of sessions &  76 & 98 & 116 & 1075 \\
Number of impressions &  280 & 297 & 471 & 3784\\
Number of $q_n\to q_{n + 1}$ pairs & 204 & 199 & 355 & 2709\\
Average number of terms in query & 3.34 & 3.40 & 3.51 & 3.21 \\
\hline
\end{tabular} \label{trec-data}
\end{table}

We conducted experiments using the TREC 2011, 2012, 2013 and 2014 Session Track data \citep{TREC:2012:Session, TREC:2013:Session}, which contains search logs collected by the TREC organizers and grouped by session. Whilst particpants were given predefined topics to search over, the organizers recorded all of the displayed URLs, titles and snippets and also user interactions including clicks and document dwell time . The corpora used were the ClueWeb09\footnotemark[\getrefnumber{clueweb9}] and ClueWeb12\footnotemark[\getrefnumber{clueweb12}] datasets. Relevance judgments were also collected for documents related to each of the topics. See Table~\ref{trec-data} for more detailed information about the datasets. 

In comparison to commercial search logs, the TREC dataset is small. Moreover, the artificial setting in which the participants were recorded conducting session search makes analysis on its data difficult to apply to commercially used search systems. For the purpose of this study, the dataset is ideal in that it is the only publicly available search log that contains the rich impression data needed for our analysis, that is, clicks, dwell times and all ranked snippets and documents (not just clicked). Whilst our statistics may not exactly reflect those found in commercial logs, we believe our theoretical insights are transferable, can be readily reproduced, and our methodology applicable to any similarly rich dataset. Furthermore, our dataset proved large enough to give us statistically significant values in our experiments. 

Sessions in the dataset are made up of a list of queries, each of which contains a ranking of $M$ documents (typically $M=10$), the snippets and titles of each document and a list of the documents that were clicked including their order and dwell time. In a session containing $N$ queries, we refer to the $n$'th query as $\qn$ and its query reformulation (if $ n < N$) as $\qr$. We denote $\overrightarrow{\qn}$ as the term vector representation of the query (with term frequency as the term weights) and $Q_n$ as the set of its terms $t_n$. 
Our analysis and experiments concern the changes between queries in a session, so we extract each pair of queries in a session $\qn \to \qr$ for $n = 1 \ldots N - 1$. 

An \textbf{Impression} refers to all of the search data related to a query such as the ranked list of documents and the clickthroughs. Elements of an impression include snippets (and their titles), clicks, dwell time and documents. In this dataset each session ends with a `test' query intentionally containing no ranking, the original purpose being for researchers to create rankings for this query by utilizing the information in the session. In these cases we do not consider this query to have an impression but we do make use of it in our query reformulation pairs unless stated otherwise. 

We used the Natural Language Toolkit (NLTK)\footnote{\url{http://www.nltk.org/}} to remove punctuation and tokenize all textual content, and then stemmed terms using the Porter Stemmer \citep{Porter:1997:ASS:275537.275705}. We opted to remove stop words, but bore in mind that this did render some query reformulations as identical to the previous query even if they originally weren't. For instance, in session 95 of the 2012 dataset, $q_1 = $`{\tt connecticut fire academy}' and  $q_2 = $`{\tt what is the connecticut fire academy}', yet after stop word removal $q_1 = q_2$. In this case, the reformulation is a more focused query than its predecessor but it nonetheless addresses the same information need with the same core terms. We used the Beautiful Soup HTML Parser\footnote{\url{http://www.crummy.com/software/BeautifulSoup/}} to extract textual content from the ClueWeb HTML documents.

We treat each term source (such as a query or snippet) as a bag of words (BoW), even though using $n$-grams could make our methodology more discernible. For example, in session 285 of the 2014 dataset, $q_1 = $`{\tt depression}' and $q_2 = $`{\tt help someone with depression}'. With BoW, we treat the terms `{\tt help}' and `{\tt someone}' separately, and we indeed find examples of the term `{\tt help}' in the snippets for $q_1$, although erroneously in the context of the web-page (`{\tt ...Help FAQ Advertising...}' at rank 3) rather than that implied by the query. Here, a bigram would distinguish `{\tt help someone}' in the correct context. Nonetheless, all of the similarity measures we use operate on a BoW model, and given that we typically only see 1 or 2 terms being added or removed from adjacent queries in a session, a unigram model is sufficient in this case. 

Our methodology concerns the analysis of text similarities. We measure the similarities of queries using the following formulae:
\begin{align}
Jaccard(Q_1, Q_2) &= \frac{|Q_1 \cap Q_2|}{|Q_1 \cup Q_2|}\\
Cosine(\overrightarrow{q_1}, \overrightarrow{q_2}) &= \frac{\overrightarrow{q_1} \cdot \overrightarrow{q_2}}{\|\overrightarrow{q_1}\|\cdot \|\overrightarrow{q_2}\|}
\end{align}
where $q_1$ and $q_2$ are queries (or any other term source). $Jaccard$ similarity is commonly used in measuring set similarity, in this case sets of terms, and $Cosine$ similarity is widely used in the vector space model in IR. 

\section{Term Retention and Removal}
\label{retention}

In our first analysis we investigate the term actions \emph{retention} and \emph{removal}. These two actions are only applied to terms found in the user's query $t_{n}$, where \emph{retention} means that $t_{n} \in Q_{n+1}$ and \emph{removal} is when $t_{n} \notin Q_{n + 1}$.

\begin{table}[t]
\centering
\setlength{\tabcolsep}{3pt}
\caption{Average number of terms retained, removed or added from $\qn \to \qr$ and the similarity between the two queries across TREC Session Track datasets.}
\begin{tabular}{r | c | c | c | c | c }
 \multicolumn{1}{ c |}{\multirow{2}{*}{ }} & \multicolumn{5}{ c}{TREC Session Track}       \\
  \cline{2-6}
   & 2011 & 2012  & 2013  & 2014 & Combined  \\
               \hline
$Jaccard(\overrightarrow{\qn}, \overrightarrow{\qr})$ & 0.49 & 0.52 & 0.50 & 0.51 & 0.50 \\
$Cosine(Q_n, Q_{n+1})$ & 0.60 & 0.65 & 0.62 & 0.63 & 0.63 \\
$\#$ terms \emph{retained} from $\qn \to \qr$ & 2.12 & 2.29 & 2.28 & 2.10 & 2.13 \\
$\#$ terms \emph{removed} from $\qn \to \qr$& 1.20 & 1.05 & 1.20 & 1.11 & 1.12 \\
$\#$ terms \emph{added}  from $\qn \to \qr$ & 1.33 & 1.35 & 1.33 & 1.21 & 1.24
\end{tabular}\label{q-qr-overlap}
\end{table}

We measured the average number of terms retained, removed or added and the average Jaccard and Cosine similarity between adjacent queries found in sessions in the TREC datasets, our results are in Table~\ref{q-qr-overlap}. We see that adjacent queries are similar to one another, with high similarity scores and term retention. We note that measures are   generally consistent across the individual datasets and their combination, and so the remainder of our analyses will be conducted on the combined dataset. We find that across all datasets, an average of 63\% of the terms in $\qr$ can be found in $\qn$, where 66\% of its terms are \emph{retained} (2.13 terms), 34\% of terms are \emph{removed} (1.12 terms) and 1.24 terms are added. 33\% of the time the reformulation contains all of the terms found in the original query. Retained terms clearly make up a large proportion of a reformulation and are indicative of the core terms defining the user's information need. 


\begin{figure*}
\centering
  \includegraphics[width=\textwidth]{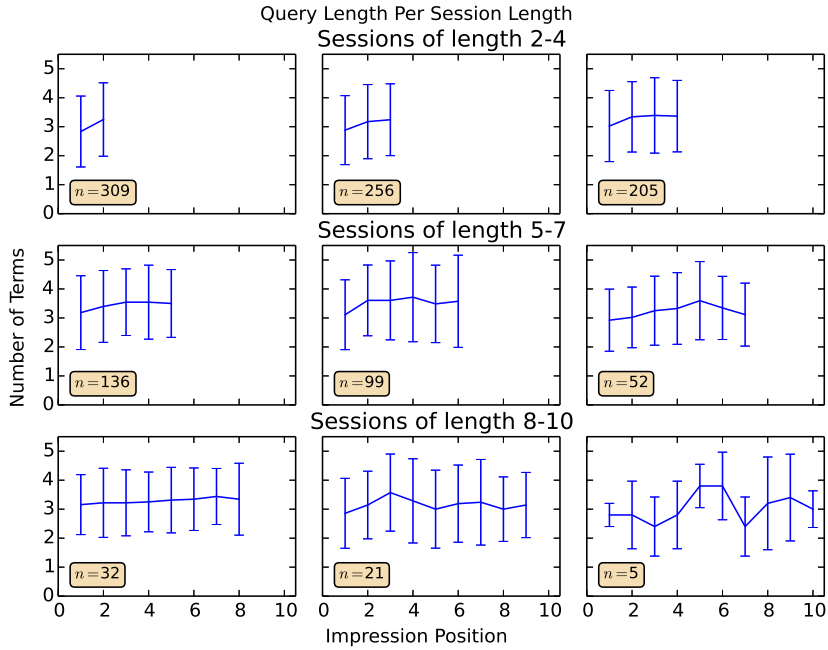}
  \caption{Plots of the average number of terms in queries at different impression positions in a session, for different lengths of session. The number of instances of each session length are labeled as $n$ in each subplot. }
\label{query-length-impression}
\end{figure*}

An important observation is that on average the length of queries increases from 3.25 terms to 3.37 terms, meaning that it cannot always be possible to source $\qr$ terms from $\qn$. To determine if this relationship holds throughout a session, we found the average query length at each impression position for a number of different session lengths (see Figure~\ref{query-length-impression}). Our results show that for shorter sessions (2 - 4 impressions) query size does appear to marginally increase, for medium session lengths (5 - 7 impressions) the query size initially increases to a point and can start to decrease, and for longer sessions (8 - 10 impressions) the query length varies unpredictably, presumably due to the small population sizes. Medium and longer sessions are likely to contain shifts in information need (for example, between queries 4 and 5 in Table~\ref{trec-example}), which may explain the variability of query length with increased impression position. It is clear from these results that reformulations can gain or lose terms depending on its position in a session. 

\begin{figure*}
\centering
  \includegraphics[width=\textwidth]{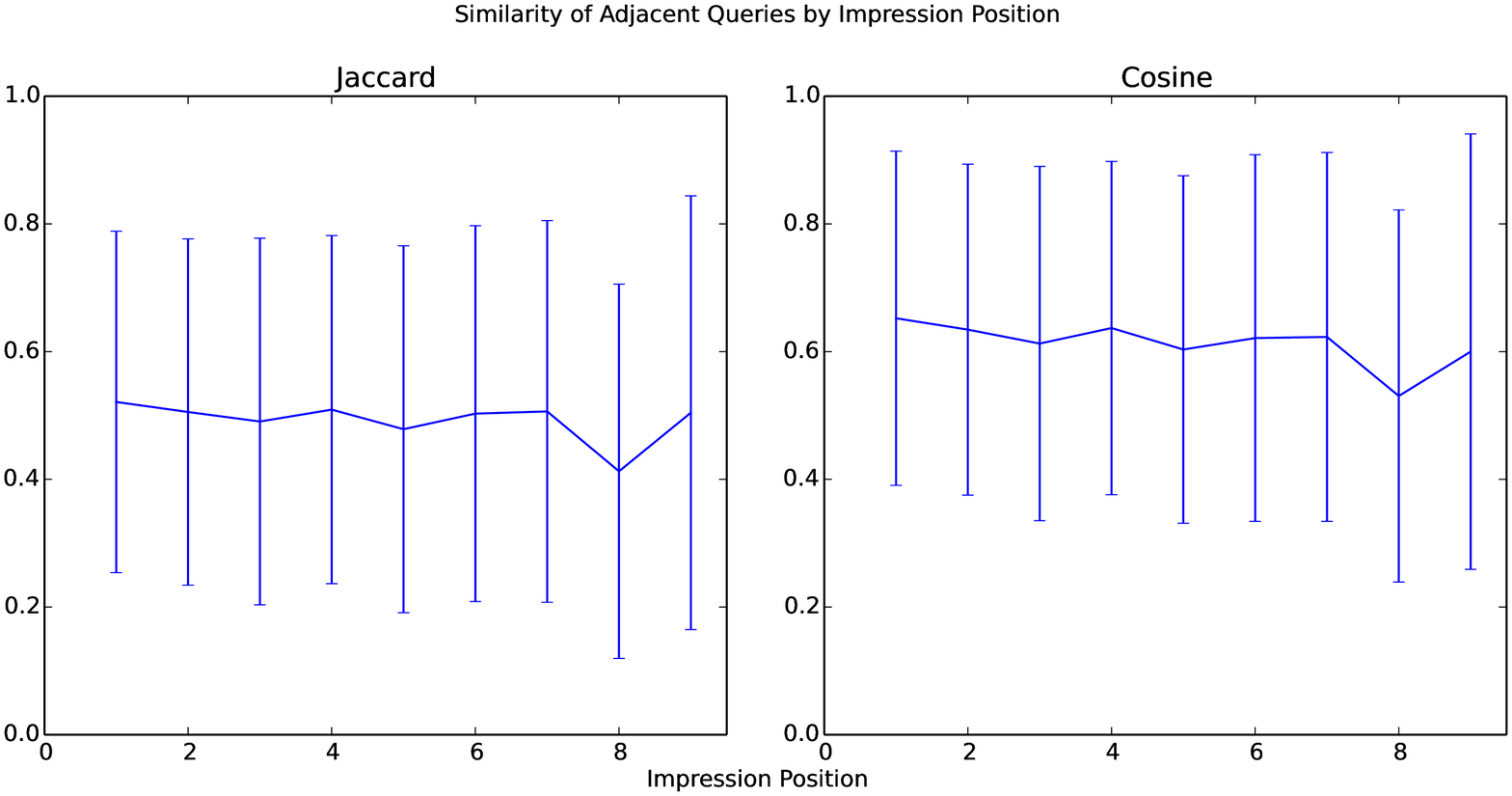}
  \caption{Average similarity of $\qn \to \qr$ pairs for impression positions $n = 1\ldots9$}
\label{overlap-per-impression}
\end{figure*}

In Figure~\ref{overlap-per-impression} we measured the similarity between query reformulations and their preceding query at each impression position. In our previous analysis we found that impression position affected query length (subject to session length), so here we investigate if this also holds for query similarity. The main conclusion we can draw is that the results are too variable to discern a pattern, with no clear trend for increasing or decreasing similarity. What this tells us is that throughout a session, queries are generally similar to their reformulations regardless of position in the session. 

Nonetheless, we do expect information needs to change throughout a session and when that happens the similarity between adjacent queries should change. For instance, in Table~\ref{trec-example} the average similarity scores between all adjacent queries are $Jaccard=0.44$ and $Cosine=0.57$, but between queries 4 and 5, the shift in query intent is captured in the change in similarity scores, calculated as $Jaccard=0.17$ and $Cosine=0.29$, a noticeable departure from the average.  

\begin{figure*}
\centering
  \includegraphics[width=\textwidth]{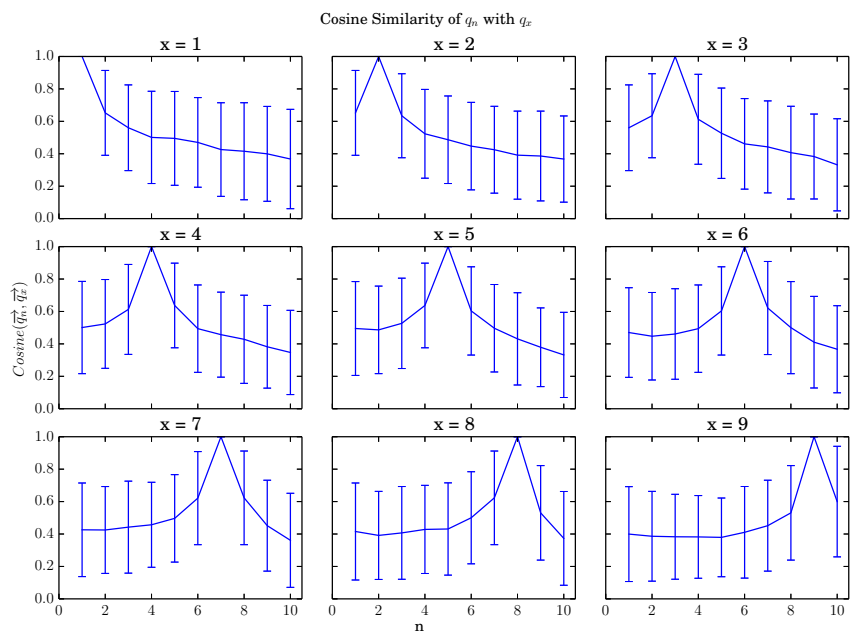}
  \caption{Cosine similarity of fixed query $q_x$ with every other query $\qn$ in the session for $x = 1\ldots9$}
\label{fixed-query-similarity}
\end{figure*}

In Figure~\ref{fixed-query-similarity} we show that core query terms do not remain constant throughout a session, indicating that the terms used in queries are always progressively changing. In this instance we picked cosine similarity although we observe the same trend for Jaccard similarity. We see that queries occurring on either side of the `fixed' query $q_x$ are the most similar but queries further away in the session become more dissimilar. This behavior holds regardless of the position of $q_x$ in the session. This and the previous result demonstrates one of the key motivations of our methodology, that there does not exist a set of `core' terms that represent the user's information need throughout the session, instead, the query and its core terms evolve as the user's information need changes. Queries at the start of a session can be very different from those at the end, and as such, term retention and removal are useful locally with adjacent queries but less so across the whole session. 

\section{Term Addition}
\label{discovery}
So far we've found that on average 63\% of the terms in query reformulations can be explained by the \emph{retained} or \emph{removed} term actions, leaving 37\% of terms unaccounted for. In this section we investigate the \emph{addition} term action which is applicable to added terms $\added$ which are terms added from $\qn$ to $\qr$ i.e. $A_{n+1} = Q_{n+1} \backslash Q_n$. Whereas before we analyzed the similarity of the query reformulation against query terms $t_{n}$, in this section we measure the similarity against terms from each of the term sources found in the impression. 

When we compare different term sources with $\added$ we run into problems caused by term source length. For instance, the $Jaccard$ similarity is sensitive to the size of the sets it compares, comparing with a larger set leads to lower similarity, making comparisons between different term sources biased. Additionally, in our studies so far we have been comparing the small number of terms found in queries, where we can consider every term important. Conversely, our term sources can contain hundreds of terms, only a few of which may match the added terms. We counteract these problems in two ways: first we use TFIDF scores \citep{SparckJones:1988:SIT:106765.106782} instead of term frequencies in our $Cosine$ similarity measure which helps us match on those added terms that are important to the term source. Thus, from this point on any term vectors $\overrightarrow{a}$ refer instead to the TFIDF vector. Secondly, we measure BM25 \citep{rz} (with typical parameter settings $k_1 = 1.2$ and $b = 0.75$) which is designed to find the similarity of queries consisting of few terms against documents with many terms, and is robust to differing document length. When we use these measures, we treat the collection of all instances of that term source as the document collection for IDF and average document length, for example, the collection of all snippets in the dataset when comparing against a snippet term source.

\subsection{Snippet Analysis}
\label{snippet-analysis}

We start by considering the snippets found in an impression. A query $q_n$ may have up to $M$ ranked snippets $s_n(k)$ where $s$ is the snippet and $k$ is its rank $1 \leq k \leq M$. In our dataset we join the snippet title onto the snippet under the assumption that anyone reading the snippet has also read its title. 

In our first study we look at the similarity of snippets $s_n$ against added terms $\added$ at different rank positions. A natural hypothesis based on eye tracking studies \cite{Granka:2004:EAU:1008992.1009079} is the concept of rank bias, that search results ranked at the top have a higher chance of being observed, thus, they should be more similar to terms added to the next query than lower ranked, potentially unobserved snippets. 

In Table~\ref{snippet-table-rank} we average similarity scores for each snippet $s_n(k)$ from rank 1 to rank $k$ in the impression. Under the assumption given by the Examination Hypothesis model \citep{clickmodels} that users examine all snippets in order from the top of the search results to the bottom, we average over \emph{all} snippets up until rank $k$, not just the snippet at that rank. Our results show that across metrics the similarity peaks at rank positions 2 and 3 before dropping with each rank. The similar lengths of snippets at each rank allows us to rule out a term source length bias. Curiously, the highest ranked snippet on its own does not have the highest similarity to added terms. The implication here is that terms used in query reformulations have a higher chance of being found in the top 2 or 3 ranked snippets and that users don't just consider the top ranked snippet on its own. We note that the examining of the top 2 or 3 search results is consistent with eye tracking observations. 

\begin{table}[t]\caption{Average similarity scores between added terms $\added$ and snippets $s_n$ up to rank $k$ in an impression. For example, if $k = 3$, then the score is the average over $s_n(1), s_n(2)$ and $s_n(3)$. Maximum values for each similarity measure are in bold.}
\centering
\begin{tabular}{r | c | c | c | c | c  }
 & \multicolumn{5}{c}{$k$} \\
 \cline{2-6}
 & 1 & 2 & 3 & 4 & 5  \\
 \hline
 $Jaccard(A_{n+1}, S_{n}(k))$ & 0.00531 & \textbf{0.00536} & 0.00529 & 0.00507 & 0.00494  \\
 $Cosine(\overrightarrow{\added}, \overrightarrow{s_n(k)})$ & 0.0184 & \textbf{0.0197} & 0.0195 & 0.0187 & 0.0185 \\
$BM25(\added, s_n(k))$ & 0.704 & 0.756 & \textbf{0.758} & 0.737 & 0.728\\
\hline
\# terms in $s_n(k)$ & 48.3 & 48.8 & 49.9 & 50.2 & 50.3 
\end{tabular}\label{snippet-table-rank}
\end{table}

From click model research we can also make the assumption that if we observe a click in an impression, then the user has examined all snippets up until that rank. Let us denote LC as the rank of the Last Click in an impression (that is, the lowest ranked clicked document). In our next study we observe whether similarity change occurs at rank LC and for the snippets ranked above and below it, akin to the `Click > No-Click Next' strategy and its variants outlined by \cite{Joachims:2005:AIC:1076034.1076063}. If an impression didn't contain a click, then we included all snippets in the impression, our results are in Table~\ref{snippet-table-lastclick}. 

We may have expected a decrease in similarity following rank LC, owing to the hypothesis that a user does not examine documents ranked lower than the last click. In our experiment we find this is not the case, recording a higher similarity score when considering all snippets in an impression rather than just up until the last clicked. A difference between our session search setting and that typically modeled with click models is that in our case, even after a document has been clicked, we know that the user returned to the set of search results in order to issue a reformulation. Conventional click models do not take into account multiple queries in a search session. As such, in our case it is likely that the user continued to examine snippets after the last click, before abandoning the query and issuing a reformulation, leading to our observed results. Also, by comparing these results with those in Table~\ref{snippet-table-rank} we see that the top ranked 2-3 snippets are still more likely to contain added terms. 

\begin{table}[t]\caption{Average similarity scores between added terms $\added$ and snippets $s_n$ up to and around rank LC in an impression, as well as all snippets. Maximum values for each similarity measure are in bold.}
\centering
\begin{tabular}{r | c | c | c | c | c  }
 & \multicolumn{5}{c}{$k$} \\
 \cline{2-6}
 & $\mbox{LC} - 1$ & $\mbox{LC}$  & $\mbox{LC} + 1$  & $\mbox{LC} + 2$  & $M$  \\
 \hline
 $Jaccard(A_{n+1}, S_{n}(k))$ & 0.00440 & 0.00446 & 0.00450 & 0.00458 & \textbf{0.00465}  \\
 $Cosine(\overrightarrow{\added}, \overrightarrow{s_n(k)})$ & 0.0167 & 0.0171 & 0.0172 & 0.0174 & \textbf{0.0175} \\
$BM25(\added, s_n(k))$ & 0.656 & 0.671 & 0.676 & 0.682 & \textbf{0.688}\\
\hline
\# terms in $s_n(k)$ & 51.0 & 51.0 & 51.0 & 50.9 & 50.4 
\end{tabular}\label{snippet-table-lastclick}
\end{table}

These inferences can be observed in our example session in Table~\ref{trec-example}. For queries $q_5 = $`{\tt gun violence us}' and $q_6 = $`{\tt law center to prevent gun violence}', which we've already noted for its shift in query intent, we observe the added term `{\tt center}' in the snippet at rank 3, which has the last (and only) clickthrough. This is in line with our findings on top ranked snippets in Table~\ref{snippet-table-rank} and corroborates our last click hypothesis. Yet, at ranks 7 and 8 we see instances of the added term `{\tt prevent}', suggesting that in this case the user examined snippets beyond the one that was clicked. 

\subsection{Term Sources}

So far we have investigated the effect of impression and rank position on similarity and introduced clicks into our last experiment. Here we directly use clicks to further distinguish between the two distinct sources of added terms in an impression, snippets and documents. This allows us to split an impression into three term sources:
\begin{description}
\item[\textbf{Non-Clicked Snippets} ($ncs$)] Snippets without a clickthrough.
\item[\textbf{Clicked Snippets} ($cs$)] Snippets with a clickthrough.
\item[\textbf{Clicked Documents} ($cd$)] Documents with a clickthrough
\end{description}

We note that the combination of $nc$ and $cs$ gives us all snippets in the impression i.e. $(\bigcup CS) \cup (\bigcup NCS) = S(M)$. We can now consider impression terms as belonging to one or more of the described term sources and start to evaluate how effective they are at providing added terms for query reformulations. Our reasoning for incorporating clicks into the term source definitions is that implicit user feedback is an indicator of the relevance of the terms contained in the source and the user's behavior at that point in the session.

\begin{table} [t]\caption{Average similarity of added terms with click-based variations of the snippet and document term sources and also the full preceding impression ($i$) and all previous impressions ($h$). Bold scores indicate a statistically significant ($p < 0.01$ under Welch's t-test) difference from non-clicked and `All' variants of the term source.} 
\centering
\begin{tabular} {r| c| c | c | c}   
Term Source & \# terms & $Jaccard$ & $Cosine$ & $BM25$ \\
\hline
All Snippets ($s(M))$ & 50.4 & 0.00465 & 0.0175 & 0.688\\
Clicked Snippets ($cs$)& 50.1 & \textbf{0.00752} & \textbf{0.0289} & \textbf{1.100}\\
Non-Clicked Snippets ($ncs$)& 50.5 & 0.00445 & 0.0167 & 0.660\\
\hline
All Documents ($ad$)& 808.8 & 0.00131 & 0.0251 & 5.612\\
Clicked Documents ($cd$)& 974.2 & \textbf{0.00171} & \textbf{0.0398} & \textbf{8.207}\\
Non-Clicked Documents ($ncd$)& 796.4 & 0.00128 & 0.0240 & 5.417\\
\hline
Impression ($i$)& 8127.2 & 0.00067 & 0.0381 & 3.535\\
Historical ($h$)& 19802.9 & 0.00052 & 0.0568 & 4.370\\
\hline
\end{tabular} \label{an-overlap}
\end{table}

Table~\ref{an-overlap} contains the results of our similarity analysis over different term sources and their variations with added terms. We compared clicked snippets and documents ($cs$ and $cd$) with their non-clicked counterparts ($ncs$ and $ncd$) and also against both combined ($s(M)$ and $ad$). We see in both cases statistically significant increases in similarity when considering clicks, a clear indicator that clicked snippets and documents are a source of terms used in query reformulations. Clicked documents score higher for the length normalized metrics $Cosine$ and $BM25$ (the score is lower for the length biased $Jaccard$ measure), indicating the importance of clicked documents. We measured the similarity of non-clicked documents in order to provide comparison with clicked documents, but ultimately we do not consider them as a term source. This is because we cannot know if the user has been exposed to them during the session, although it is feasible that the user has encountered the document before or was satisfied by the non-clicked snippet itself. 

We also measured the similarity with all terms found in the impression, where $I = S(M) + CD$ (not including the query). We find that differentiating an impression into click based term sources does lead to improved similarity scores. Taking this further, we also measured against historical impressions, i.e. all impression terms that occur earlier in the session up to and including the current impression $H_n = \bigcup_{j = 1}^n I_j$, to test the assumption that users obtain terms not just from the preceding impression but also those encountered earlier. For instance, in our example in Table~\ref{trec-example}, the term `{\tt current}' from $q_3$ is not found in the preceding impression for $q_2$, whereas it occurs 3 times in the snippet at rank 3 of $q_1$.  We do see an increase in similarity scores over the historical impression terms and values that are comparable with the other term sources, suggesting that terms can be sourced from earlier in the session. In this work we define our term sources based only on the preceding impression, but using earlier impressions could prove an interesting extension. 

\subsection{Dwell Time}

From Table~\ref{an-overlap} we see that clicked documents have substantially more terms than snippets. A central argument of our methodology is that users choose reformulation terms that they have been exposed to from term sources, hence, in order to come across terms in a long document, time must be spent reading it. Our dataset records the dwell time of each clicked document, which is an indicator of reading time.

We find that the average dwell time is 35.3 seconds before users return to the set of search results. This is similar to the 30 second threshold used in other IR research as a marker for a satisfactorily (SAT) clicked document \citep{White:2007:IBV:1242572.1242576}. SAT clicks are often used as a replacement for relevance judgments in the absence of human assessors, usually on large query logs. We find that a dwell time threshold of 30 seconds differentiates 40\% of the clicked documents. 

\begin{figure*}
\centering
  \includegraphics[width=0.8\textwidth]{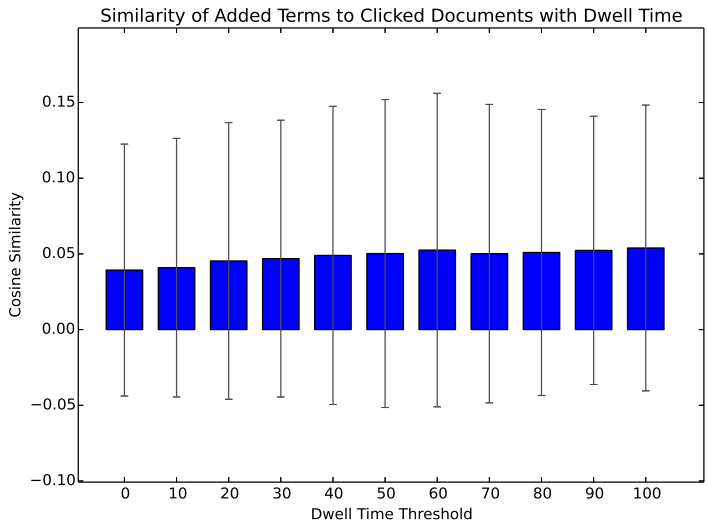}
  \caption{Average $Cosine$ similarity of added terms with clicked documents at different dwell time threshold levels. }
  \label{dwell-time}
\end{figure*}

To test whether dwell time should be considered a feature in our methodology, we measured the similarity of clicked documents against added terms at a range of different dwell time thresholds. Figure~\ref{dwell-time} displays the results for $Cosine$ similarity, the other measures reported similar findings. Whilst we do observe a slight increase in similarity with dwell time threshold, the results are too variable to be able to draw any conclusions. In particular, the SAT click threshold does not appear to offer any clear indicator of improvement. Our findings are supported by recent research that argues that this single value cannot capture the complexities of reading behavior and user satisfaction \citep{Kim:2014:MDT:2556195.2556220}. As such, we do not consider dwell time as a feature in our methodology and instead use all clicked documents as a term source collectively.

\section{Term Scenario Analysis}
\label{experiment}

Our term-based methodology has given us insight into the circumstances where terms are retained, removed or added to query reformulations. Use of the similarity measures has helped us define the three term sources, based on user interactions, that influence the terms added to the next query in a session. In this section, we extend our methodology to measure how effective query reformulations are under different circumstances. We do this by defining 8 user behavior scenarios based on the combination of term sources, which can help interpret our results and understand user motivations. 

\subsection{Query and Added Term Scenarios}

We first focus on the query terms $t_{n}$ and whether the term actions \emph{retention} or \emph{removal} are usually applied to them by the user. To expand on the limited information available to us on the terms in the query, we can look for occurrences of the term in the impression. More specifically, the three term sources $ncs$, $cs$ and $cd$. $t_{n}$ can belong to any combination of term sources, including all or none, giving 8 query \textbf{term scenarios}. Each combination of term source defines a scenario and we give a full index of scenario number definitions in Table~\ref{scenarios}. 

In our previous analysis we were able to make inferences on terms based on which term source they originated from. With the expansion of 3 term sources to 8 scenarios we can now make more interesting observations. For instance, in the first query in our example (Table~\ref{trec-example}), the terms `{\tt gun}' and `{\tt control}' both belong to scenario 8 (they appear in non-clicked and clicked snippets and also clicked documents) and they are retained in the query reformulation. Conversely, the term `{\tt opinions}' is only found in non-clicked snippets (scenario 5) and is subsequently removed. An inference we can make here is that finding query terms in clicked snippets and documents is a strong indicator that the term will be kept, whereas query terms that only appear in non-clicked snippets are more likely to be removed.

We also assign added terms to the same scenarios in Table~\ref{scenarios}. Given that the purpose of our methodology is to understand when terms from the previous impression (including query) will be used in the reformulation, we appreciate that a real search system would not have access to added terms in order to assign them to scenarios. Nonetheless, by analyzing these terms in the same way as query terms, we gain insight into which circumstances a user is likely to add terms from the impression. 

\begin{table} [t]\caption{Overall average clicks, non-clicks and documents per impression and overall number of query and added term scenarios.}
\label{overall-data}
\centering
\begin{tabular} {  r | c } 
\# ranked documents	& 10.5\\
\# clicks	& 0.626\\
\# non-clicks & 9.87\\
\hline
\# query term scenarios & 7621\\
\# added term scenarios &	2981 \\
\end{tabular} \label{overall-data}
\end{table}

We extracted all query reformulation pairs from the dataset as before but this time did not include test queries (the final query in each session). Test queries do not contain rankings or relevance judgments, and thus are unsuitable for our evaluations in the next subsections. We assigned terms from $q_n$ and $\added$ to each scenario and give an overview of our results in Tables~\ref{overall-data} and \ref{scenarios}.

\begin{table} [t]\caption{Scenario number definitions, occurrence \% for query and added term scenarios and average number of ranked documents and clicks for each scenario.}
\centering
\begin{tabular} {c | c | c | c | c | c | c | c | c | c} 
\multicolumn{1}{ c |}{\multirow{2}{*}{Scenario}} & \multicolumn{3}{c |}{$t \in$} & \multicolumn{3}{c | }{Query term scenarios} & \multicolumn{3}{c }{Added term scenarios}\\
 \cline{2-10}
  & \multicolumn{1}{ c |}{$ncs$} & $cs$ & $cd$ & \% & \# Docs & \# Clicks &  \% & \# Docs & \# Clicks\\
\hline
1 & False & False & False & 9.95 & 8.64 & 0.27 & 57.0 & 10.2 & 0.38 \\
2 & False & False & \cellcolor{blue!25}True & 0.35 & 11.1 & 1.89 & 7.85 & 10.4 & 1.86\\
3 & False & \cellcolor{blue!25}True & False & 0.05 & 3.50 & 1.25 & 0.20 & 7.00 & 1.00\\
4 & False & \cellcolor{blue!25}True & \cellcolor{blue!25}True & 0.68 & 10.2 & 2.27 & 2.01 & 11.0 & 2.17\\
5 & \cellcolor{blue!25}True & False & False & 60.2 & 10.5 & 0.06 & 24.2 & 10.8 & 0.15\\
6 & \cellcolor{blue!25}True & False & \cellcolor{blue!25}True & 2.27 & 10.6 & 1.41 & 4.43 & 10.4 & 1.46\\
7 & \cellcolor{blue!25}True & \cellcolor{blue!25}True & False & 0.42 & 11.7 & 0.94 & 0.07 & 10.0 & 0.50\\
8 & \cellcolor{blue!25}True & \cellcolor{blue!25}True & \cellcolor{blue!25}True & 26.1 & 10.9 & 1.70 & 4.26 & 11.7 & 1.91\\
\hline
\end{tabular} \label{scenarios}
\end{table}

In Table~\ref{scenarios} we see that both sets of term scenarios are variably distributed. Scenario 5, which refers to the case where terms only appear in non-clicked snippets, is the most common scenario for query terms, comprising 60.2\% of the data. For this scenario we find that the average number of clicks is 0.06, well below the overall average in Table~\ref{overall-data}. Thus, scenario 5 appears to be capturing the common case where users do not click on any results, hence no other term source matching occurs. Scenario 8 makes up a further 26.1\% of cases and represents the situation where terms appear in all term sources. We would expect query terms to appear in snippets (either $ncs$ or $cs$) and we find that this is the case 90\% of the time. Interestingly, 9.95\% of query terms do not appear in the impression at all.

We see a different distribution of scenarios for added terms, the most prominent being scenario 1 at 57\%. This is the case where added terms cannot be found in the previous impression and mirrors the findings in Table~\ref{q-qr-overlap}. Scenario 5 is also common for added terms. The four scenarios with terms found in clicked documents (2, 4, 6 and 8) make up 18.6\% of the scenarios, further evidence of clicked documents being a valuable source of added terms. There is a noticeable difference in occurrences between query and added terms in scenarios 2 and 4. Scenarios 3 and 7 rarely appear for both query and added terms, this can be explained by the fact that these are the cases where terms appear in clicked snippets but not clicked documents. Given that the snippet is derived from the document itself, this makes it unlikely for these scenarios to occur, and we ignore them in future analyses. 

\subsection{Term Actions}

\begin{figure*}
\centering
  \includegraphics[width=0.8\textwidth]{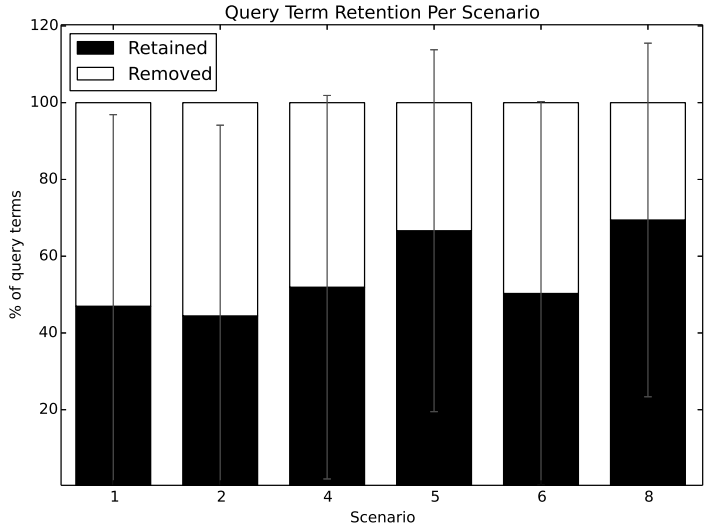}
  \caption{Proportion of query terms that are retained or removed per term scenario}
\label{scenario-term-retention}
\end{figure*}

Query term scenarios fall into two term action categories, \emph{retained} or \emph{removed}. Figure~\ref{scenario-term-retention} shows the proportion of query terms that are retained or removed from query reformulations for each scenario. Our first observation is that the two most common scenarios (5 and 8) lead to high term retention rates that are around the overall average term retention of 66\%. This coincides with our earlier finding that users generally retain terms between adjacent queries, thus, the core terms are falling into these scenario numbers. For example, in Table~\ref{trec-example} the query terms `{\tt gun control}' both belong to scenario 8 for the first 2 queries and are retained. For queries 3 and 4 (which are identical), they change to scenario 5 and are then subsequently removed in the next query. 

Scenarios 2, 4 and 6, which capture instances of query terms appearing in clicked documents, occur infrequently for query terms and here seem to lead to the removal of terms. One inference is that terms appearing in clicked documents may be removed in lieu of the user satisfying that particular search intent. We also see low retention for query terms that are not found in the impression at all, potentially an indication that the term was not useful in helping the user's search. 

\subsection{Term Scenario Evaluation}

So far we've sought to understand the term actions retention, removal and addition without explicitly evaluating whether or not they are beneficial. Simply determining which terms from queries and term sources are likely to appear in a reformulation, based on user behavior in search logs, does not necessarily mean that they will improve the search experience. These evaluations demonstrate that our methodology is able to differentiate scenarios which may lead to future clicks or improvements in IR metric scores. 
		
\subsubsection{Click Based Evaluation}

\begin{table} [t]\caption{Percentage of term scenarios and term actions that led to a click in the next query}
\centering
\begin{tabular} {  c | c | c | c } 
\multirow{2}{*}{Scenario} & \multicolumn{3}{c}{\% (Term action $\to$ Click)}\\
\cline{2-4}
  & Retained & Removed & Added \\
\hline
1 & 22.5 & 29.1 & 26.3 \\
2 & 25.0 & 53.3 & 54.3\\
4 & 59.3 & 68.0 & 63.3\\
5 & 24.5 & 22.1 & 27.5\\
6 & 41.4 & 52.3 & 59.1\\
8 & 52.3 & 49.1 & 64.6\\
\hline
\end{tabular} \label{sat-eval}
\end{table}

Our first evaluation method involves observing whether the next impression in the session contains a click, an implicit measure of success and one tied to the user whose session we are analyzing. In this experiment, for each term scenario we measured the proportion of times each of the three term actions (retaining, removing or adding) led to a click in the next impression and give our results in Table~\ref{sat-eval}. Firstly, we find that all term actions in scenarios 1 and 5 (where terms are not found in clicked snippets or documents) are less likely to lead to a click. When clicked documents are taken into account (scenarios 2, 4, 6 and 8) the likelihood of a click in the next query is much higher. In particular, for scenarios 2 and 4 clicks were more likely after removing query terms then retaining them, a result mirroring what we found in Figure~\ref{scenario-term-retention}. Terms added from clicked documents and snippets were also highly likely to result in a click. 

\subsubsection{IR Metric Based Evaluation}

\begin{table} [t]\caption{Change in value for metrics NDCG, NERR and MAP from $q_n \to q_{n+1}$ for each term action and term scenario. Bold values indicate a statistically significant difference in IR metric score ($p < 0.05$ under the Wilcoxon signed rank test).}
\centering
\begin{tabular} {  c | c| c | c | c | c | c | c} 
\multirow{2}{*}{} & \multirow{2}{*}{} & \multicolumn{6}{c}{Scenario}\\
\cline{3-8}
 &  & 1 & 2 & 4 & 5 & 6 & 8\\
 \hline
\multirow{3}{*}{Retained}  & NDCG & $\textbf{0.078} \blacktriangle$& $-0.205 \blacktriangledown$&$ \textbf{-0.120} \blacktriangledown$& $-0.009 \blacktriangledown$& $-0.019 \blacktriangledown$ &$ \textbf{-0.058}\blacktriangledown$\\
 & NERR & $\textbf{0.080} \blacktriangle$& $-0.216 \blacktriangledown$&$ \textbf{-0.146}\blacktriangledown$ & $-0.004\blacktriangledown$ & $-0.024\blacktriangledown$ & $\textbf{-0.064}\blacktriangledown$\\
 & MAP & $\textbf{0.004} \blacktriangle$& $\textbf{-0.011}\blacktriangledown$ & $0.010 \blacktriangle$& $0.001 \blacktriangle$& $-0.003\blacktriangledown$ &$ \textbf{-0.009}\blacktriangledown$\\
 \hline
\multirow{3}{*}{Removed}  & NDCG & $\textbf{0.058}\blacktriangle$ & $0.000 $& $-0.069\blacktriangledown$ &$ 0.006\blacktriangle$ &$ 0.006\blacktriangle $&$ \textbf{-0.148}\blacktriangledown$\\
 & NERR & $\textbf{0.037}\blacktriangle$ &$ -0.024\blacktriangledown$ &$ -0.063\blacktriangledown$ &$ -0.015\blacktriangledown $& $0.017\blacktriangle $& $\textbf{-0.140}\blacktriangledown$\\
 & MAP &$ \textbf{0.005} \blacktriangle$&$ -0.004\blacktriangledown$ & $0.000 $&$ \textbf{-0.003} \blacktriangledown$&$ 0.001 \blacktriangle$&$ \textbf{-0.010}\blacktriangledown$\\
  \hline
\multirow{3}{*}{Added}  & NDCG &$ -0.025 \blacktriangledown$&$ \textbf{-0.127} \blacktriangledown$&$ -0.051\blacktriangledown$ &$ -0.023\blacktriangledown$ &$ -0.046\blacktriangledown$ & $-0.091\blacktriangledown$\\
 & NERR &$ \textbf{-0.019} \blacktriangledown$&$ \textbf{-0.123}\blacktriangledown$ &$ -0.082\blacktriangledown$ &$ -0.020 \blacktriangledown$&$ -0.052 \blacktriangledown$& $\textbf{-0.073}\blacktriangledown$\\
 & MAP &$\textbf{-0.007}\blacktriangledown $&$ \textbf{-0.007}\blacktriangledown$ &$ 0.002 \blacktriangle$&$ 0.000 $&$ -0.006\blacktriangledown$ &$ -0.006\blacktriangledown$\\
\hline
\end{tabular} \label{ir-metric-eval}
\end{table}

Whilst clicks are important implicit signals of relevance, we can also make use of the TREC Session Track relevance judgments to evaluate the effectiveness of term actions. The majority of sessions in the dataset are linked to topics, for which documents have been assessed for relevance by human assessors on a scale from 0 to 4. For each impression in the data set we calculated the Normalized Expected Reciprocal Rank  at rank position 10 (NERR), Normalized Discounted Cumulative Gain at position 10 (NDCG) and the Mean Average Precision (MAP). These metrics are widely used and well regarded in the IR community and the cutoff point at rank 10 was chosen in order to evaluate the quality of results in a typical impression. NERR is a metric that rewards displaying a highly relevant document at a high rank, NDCG measures the quality of the retrieved documents and their order and MAP balances precision and recall. 

\begin{figure*}
\centering
  \includegraphics[width=0.8\textwidth]{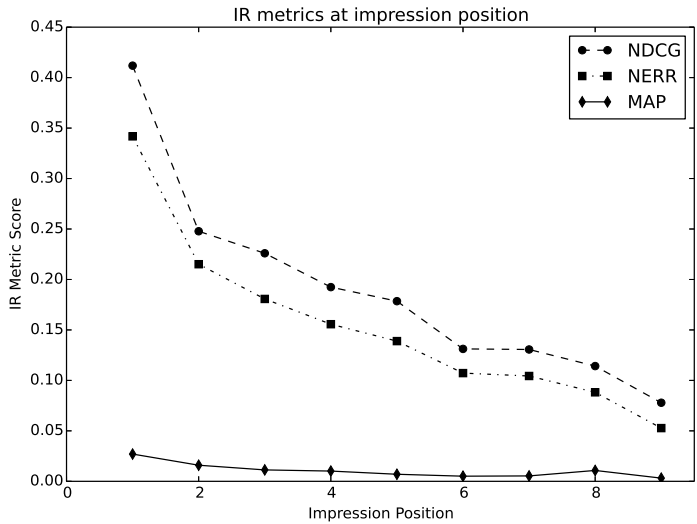}
  \caption{NERR@10, NDCG@10 and MAP scores for user created rankings at each impression position.}
\label{ir-metric-impressions}
\end{figure*}

We measured the difference in scores for each of the metrics calculated for the rankings of $\qn$ and $\qr$ across each scenario and term action and our results are in Table~\ref{ir-metric-eval}. We see that when scenario 1 query terms are \emph{retained} there is a significant improvement across all IR metrics, but otherwise for the other scenarios we see scores decreasing, significantly so for scenario 8. We also see a similar pattern for \emph{removing} terms across all scenarios. Finally, for \emph{added} terms the IR metrics decrease across all scenarios, significantly so for scenarios 1 and 2. These results indicate the existence of a general trend of decreasing IR score for adjacent queries, and we find that when we plot the scores across impression positions (Figure~\ref{ir-metric-impressions}) we confirm this negative trend. What we can take from these results is that when we come across an impression which doesn't contain query terms, the next query is likely to be an improvement (regardless of if the query term is retained or removed). Furthermore, in the converse scenario where query terms appear in all term sources, the next search ranking is likely to be worse. 

We conclude on an interesting final result, where we see that when a term is added from a clicked document only (scenario 2), it leads to rankings with poorer IR scores. This is in spite of many of our findings that indicate that clicked documents are a rich source of added terms, that scenario 2 commonly occurs and that such reformulations lead to clicks 54.3\% of the time. For example, the terms `{\tt us government}' in Table~\ref{trec-example} fall into scenario 2 for $q_1$ and are then added to $q_2$, whose ranking leads to a click and an improvement in NERR and NDCG, but not MAP, and then they are removed. Here, these terms represented a subtopic in the user's overall information need that was satisfied by their results before moving on. This result supports our argument that simply following the query reformulation behavior of users does not necessarily lead to improved search systems, but through understanding the interactions with our methodology we can make more informed inferences.

\section{Discussion}
\label{discussion}

Our novel methodology and term-based approach to understanding query reformulation leads to some interesting as well as expected results. We confirm that a user's query reformulation is largely made up of terms retained from their preceding query, with the remainder made up of a mix of terms discovered in the impression and externally sourced terms, although this can fluctuate throughout a session.  However, we cannot expect to find all terms in $\added$ based on what's available in a query log because users introduce terms based on their own cognitive processes, memory, external context or when changing their information need. For instance, in the example in Table~\ref{trec-example}, the final query contains the term `{\tt law}'  that isn't found in any of the term sources in the previous impression and it's clear from the table that this query is a departure from the topic and pattern of the previous queries. In such cases, techniques such as behavioral modeling, ontologies, contextual retrieval and topic modeling could be used to predict new terms to add but this is beyond the scope of this work.

This work could be extended by further breaking down an impression into new term sources, such as snippet and document title or document components such as headers and paragraph text. Features such as rank and impression position or click order could be used to separate the current term sources and increase the number of scenarios. An $n$-gram model would require different similarity measures but would allow more accurate phrase matching and new term actions (such as phrase rearrangement, splitting etc.). Term sources from non-adjacent impressions could also help improve the overall model, and other implicit user measures (such as mouse tracking or reading level) could prove a good differentiator of term source similarity. 

We are also aware that our analysis is restricted by the size and nature of the TREC session track data. An ideal analysis would be conducted over commercial query logs but these are not readily available. Also, the TREC data is flawed in that it has been compiled by researchers and doesn't strictly reflect an actual user interacting with a search engine. Nonetheless, the data does make up for these shortcomings with its rich meta-data, standardization and availability. Our inferences on query reformulation understanding are transferable to other areas of IR and our methodology can be readily applied to other datasets. 

Our evidence suggests that user created query reformulations are not always successful and that it may be possible to generate viable reformulations (or suggestions) based on observing user feedback and classifying which scenarios terms belong to. Our intention is to use this research to build a query suggestion agent based on a Markov Decision Process (MDP) that incorporates our methodology, allowing us to create ranked lists of query suggestions using the retention, removal and addition term actions.  By modeling the user's feedback using a Dynamic IR model \citep{Jin:2013:IES:2488388.2488446}, we can optimize the MDP over several projected queries in the session and predict the changing queries of the user, which will let us rank the most optimal query suggestion. 

\section{Conclusion}
\label{conclusion}

We have introduced a methodology for interpreting query reformulation behavior based around the three term actions \emph{retention}, \emph{removal} and \emph{addition}. We directly applied our technique in an empirical analysis over TREC Session Track Data where we analyzed the origin of terms used in query reformulations. We identified the preceding query as the main source but also found that terms located in the impression itself were an additional source. We found that adjacent queries in session tended to be very similar but that there often isn't a set of core teams that are used throughout, instead the core teams change in the session as the information need changes. 

We tested our methodology on well understood findings in click model research and found evidence of rank bias affecting reformulation behavior. We identified three user interaction based sources of terms (and discarded another based on dwell time) that are found in each impression and we tested from which sources users were able to locate terms to add to query reformulations. By matching query and impression terms in the term sources we defined a number of possible user behavior scenarios that a term could belong to. 

We measured the effectiveness of the term actions per scenario to evaluate how good they were at not just predicting query reformulations, but effective ones. By interpreting the behavior of the user for given scenarios and their corresponding effective actions, we are able to understand a user's motivations for retaining, removing or adding terms. As future work, we can make inferences and predictions of evolving queries in session search leading to better query suggestion agents, user behavior models and more accurate click log mining.


\bibliographystyle{spbasic}      
\bibliography{refs}   

\begin{thebibliography}{26}
\providecommand{\natexlab}[1]{#1}
\providecommand{\url}[1]{{#1}}
\providecommand{\urlprefix}{URL }
\expandafter\ifx\csname urlstyle\endcsname\relax
  \providecommand{\doi}[1]{DOI~\discretionary{}{}{}#1}\else
  \providecommand{\doi}{DOI~\discretionary{}{}{}\begingroup
  \urlstyle{rm}\Url}\fi
\providecommand{\eprint}[2][]{\url{#2}}

\bibitem[{Cole et~al(2010)Cole, Gwizdka, Bierig, Belkin, Liu, Liu, and
  Zhang}]{Cole:2010:LST:1962300.1962323}
Cole MJ, Gwizdka J, Bierig R, Belkin NJ, Liu J, Liu C, Zhang X (2010) Linking
  search tasks with low-level eye movement patterns. ACM, ECCE '10, pp 109--116

\bibitem[{Cole et~al(2011)Cole, Gwizdka, Liu, Bierig, Belkin, and
  Zhang}]{Cole2011346}
Cole MJ, Gwizdka J, Liu C, Bierig R, Belkin NJ, Zhang X (2011) Task and user
  effects on reading patterns in information search. Interacting with Computers
  23(4):346 -- 362

\bibitem[{Craswell et~al(2008)Craswell, Zoeter, Taylor, and
  Ramsey}]{clickmodels}
Craswell N, Zoeter O, Taylor M, Ramsey B (2008) An experimental comparison of
  click position-bias models. ACM, WSDM '08, pp 87--94

\bibitem[{Granka et~al(2004)Granka, Joachims, and
  Gay}]{Granka:2004:EAU:1008992.1009079}
Granka LA, Joachims T, Gay G (2004) Eye-tracking analysis of user behavior in
  www search. ACM, SIGIR '04, pp 478--479

\bibitem[{Guan et~al(2013)Guan, Zhang, and
  Yang}]{Guan:2013:UQC:2484028.2484055}
Guan D, Zhang S, Yang H (2013) Utilizing query change for session search. ACM,
  SIGIR '13, pp 453--462

\bibitem[{Huang and Efthimiadis(2009)}]{Huang:2009:AEQ:1645953.1645966}
Huang J, Efthimiadis EN (2009) Analyzing and evaluating query reformulation
  strategies in web search logs. ACM, CIKM '09, pp 77--86

\bibitem[{Jansen et~al(2005)Jansen, Spink, and
  Pedersen}]{Jansen:2005:TCA:1059467.1059470}
Jansen BJ, Spink A, Pedersen J (2005) A temporal comparison of altavista web
  searching: Research articles. J Am Soc Inf Sci Technol 56(6):559--570

\bibitem[{Jansen et~al(2009)Jansen, Booth, and
  Spink}]{Jansen:2009:PQR:1568763.1568772}
Jansen BJ, Booth DL, Spink A (2009) Patterns of query reformulation during web
  searching. J Am Soc Inf Sci Technol 60(7):1358--1371

\bibitem[{Jiang et~al(2014)Jiang, He, and
  Allan}]{Jiang:2014:SBC:2600428.2609633}
Jiang J, He D, Allan J (2014) Searching, browsing, and clicking in a search
  session: Changes in user behavior by task and over time. ACM, SIGIR '14, pp
  607--616

\bibitem[{Jin et~al(2013)Jin, Sloan, and Wang}]{Jin:2013:IES:2488388.2488446}
Jin X, Sloan M, Wang J (2013) Interactive exploratory search for multi page
  search results. In: WWW '13

\bibitem[{Joachims et~al(2005)Joachims, Granka, Pan, Hembrooke, and
  Gay}]{Joachims:2005:AIC:1076034.1076063}
Joachims T, Granka L, Pan B, Hembrooke H, Gay G (2005) Accurately interpreting
  clickthrough data as implicit feedback. ACM, SIGIR '05, pp 154--161

\bibitem[{Kanoulas et~al(2011)Kanoulas, Carterette, Hall, Clough, and
  Sanderson}]{TREC:2011:Session}
Kanoulas E, Carterette B, Hall M, Clough P, Sanderson M (2011) Overview of the
  trec 2011 session track. In: TREC'11

\bibitem[{Kanoulas et~al(2012)Kanoulas, Carterette, Hall, Clough, and
  Sanderson}]{TREC:2012:Session}
Kanoulas E, Carterette B, Hall M, Clough P, Sanderson M (2012) Overview of the
  trec 2012 session track. In: TREC'12

\bibitem[{Kanoulas et~al(2013)Kanoulas, Carterette, Hall, Clough, and
  Sanderson}]{TREC:2013:Session}
Kanoulas E, Carterette B, Hall M, Clough P, Sanderson M (2013) Overview of the
  trec 2013 session track. In: TREC'13

\bibitem[{Kim et~al(2014)Kim, Hassan, White, and
  Zitouni}]{Kim:2014:MDT:2556195.2556220}
Kim Y, Hassan A, White RW, Zitouni I (2014) Modeling dwell time to predict
  click-level satisfaction. ACM, WSDM '14, pp 193--202

\bibitem[{Kinley et~al(2012)Kinley, Tjondronegoro, Partridge, and
  Edwards}]{Kinley:2012:HII:2414536.2414586}
Kinley K, Tjondronegoro D, Partridge H, Edwards S (2012) Human-computer
  interaction: The impact of users' cognitive styles on query reformulation
  behaviour during web searching. ACM, OzCHI '12, pp 299--307

\bibitem[{Liu et~al(2010)Liu, Gwizdka, Liu, Xu, and Belkin}]{ChangLiu:2010}
Liu C, Gwizdka J, Liu J, Xu T, Belkin NJ (2010) Analysis and evaluation of
  query reformulations in different task types. In: ASIST '10

\bibitem[{Liu and Belkin(2010)}]{Liu:2010:PIR:1835449.1835457}
Liu J, Belkin NJ (2010) Personalizing information retrieval for multi-session
  tasks: The roles of task stage and task type. In: SIGIR '10

\bibitem[{Liu et~al(2011)Liu, Miao, Zhang, Ma, and Ru}]{Liu201113847}
Liu Y, Miao J, Zhang M, Ma S, Ru L (2011) How do users describe their
  information need: Query recommendation based on snippet click model. Expert
  Systems with Applications 38(11):13,847 -- 13,856

\bibitem[{Liu et~al(2014)Liu, Wang, Zhou, Nie, Zhang, and
  Ma}]{Liu:2014:SRT:2661829.2661907}
Liu Y, Wang C, Zhou K, Nie J, Zhang M, Ma S (2014) From skimming to reading: A
  two-stage examination model for web search. ACM, CIKM '14, pp 849--858

\bibitem[{Marchionini(2006)}]{Marchionini:2006:ESF:1121949.1121979}
Marchionini G (2006) Exploratory search: From finding to understanding. Commun
  ACM 49(4):41--46

\bibitem[{Porter(1997)}]{Porter:1997:ASS:275537.275705}
Porter MF (1997) Readings in information retrieval. Morgan Kaufmann Publishers
  Inc., chap An Algorithm for Suffix Stripping, pp 313--316

\bibitem[{Robertson and Zaragoza(2009)}]{rz}
Robertson S, Zaragoza H (2009) The probabilistic relevance framework: Bm25 and
  beyond. Found Trends Inf Retr 3(4):333--389

\bibitem[{Song et~al(2009)Song, Luo, Nie, Yu, and
  Hon}]{Song:2009:IAQ:1508326.1508674}
Song R, Luo Z, Nie JY, Yu Y, Hon HW (2009) Identification of ambiguous queries
  in web search. Inf Process Manage 45(2):216--229

\bibitem[{Sparck~Jones(1988)}]{SparckJones:1988:SIT:106765.106782}
Sparck~Jones K (1988) Document retrieval systems. Taylor Graham Publishing,
  chap A Statistical Interpretation of Term Specificity and Its Application in
  Retrieval, pp 132--142

\bibitem[{White and Drucker(2007)}]{White:2007:IBV:1242572.1242576}
White RW, Drucker SM (2007) Investigating behavioral variability in web search.
  WWW '07, pp 21--30

\end{thebibliography}


\end{document}